\documentclass[aps,prb,twocolumn,showpacs]{revtex4-1}
\usepackage{amssymb,amsmath}
\usepackage{graphicx}
\usepackage{dcolumn}
\usepackage{bm}
\usepackage{color}

\begin{document}

\title{Dynamic structure factor of liquid $^{\bf 4}$He across the
normal-superfluid transition}

\author{G. Ferr\'e and J. Boronat}

\email{jordi.boronat@upc.edu}
\affiliation{Departament de F\'{i}sica, Universitat Polit\`{e}cnica
de Catalunya, Campus Nord B4-B5, E-08034, Barcelona, Spain}

\begin{abstract}
We have carried out a microscopic study of the dynamic structure factor of
liquid $^4$He across the normal-superfluid transition temperature using the
path integral Monte Carlo method. The ill-posed problem of the inverse
Laplace transform, from the imaginary-time intermediate scattering function
to the dynamic response, is tackled by stochastic optimization. Our results
show a quasi-particle peak and a small and broad multiphonon contribution.
In spite of the lack of strength in the collective peaks, we clearly
identify the rapid dropping of the roton peak amplitude 
when crossing the transition
temperature $T_\lambda$. Other properties such as the static structure factor,
static response, and one-phonon contribution to the response are also
calculated at different temperatures. The changes of the phonon-roton
spectrum with the temperature are also studied. An overall agreement with
available experimental data is achieved.

\end{abstract}

\pacs{67.25.dt,02.70.Ss,02.30.Zz }
\maketitle

\section{Introduction}
\label{sec:introduction}

The most relevant information on the dynamics of a quantum liquid is
contained in the dynamic structure factor $S({\bm q},\omega)$, which is
experimentally measured by means of inelastic neutron
scattering.~\cite{lovesey} Probably,
superfluid $^4$He has been the most deeply studied system from both theory
and experiment and a great deal of information about it is nowadays
accessible.~\cite{glyde} For many years, liquid $^4$He was the only quantum fluid showing
Bose-Einstein condensation and superfluidity until the discovery of the
fully Bose-Einstein condensate in 1995.~\cite{cornell,ketterle} Therefore, 
the number of measures
of $S({\bm q},\omega)$ at different temperatures and momentum transfer has
been continuously growing, with more refined data along the
years.~\cite{andersen,fak,glyde2,gibbs,pearce,sakhel} The
emergence of strong quasi-particle peaks going down the normal-superfluid
transition ( $T_\lambda=2.17$ K) has been associated with the
superfluidity of the system by application of the Landau criterium. Much of
the interest on the dynamics of strongly-correlated liquid $^4$He is
then related to the effects on the dynamics of this second-order
$\lambda$-transition.  

In the limit of zero temperature, the richest and most accurate microscopic
description of the dynamic response of liquid $^4$He has been achieved by
progressively more sophisticated correlated basis function (CBF)
theory.~\cite{fabro}
The development of this theory has been stimulated by the continuous
improvement of the experimental resolution in inelastic neutron scattering.
Recently, Campbell \textit{et al.}~\cite{campbell} have incorporated 
three-body fluctuations in 
an extended CBF approach and proved a remarkable improvement of both the
excitation spectrum and full $S({\bm q},\omega)$, with features not so
clearly seen before and that are in nice agreement with the most recent
experimental data.~\cite{godfrin} On the other hand, the most accurate tools to deal with
ground-state properties are the quantum Monte Carlo (QMC) methods. In the case of
bosons as $^4$He, these methods are able to produce essentially exact
results for its equation of state and structure properties which are in
close agreement with experimental data.~\cite{borobook} Importantly, QMC methods are not
restricted to the limit of zero temperature and are equally powerful to
introduce the temperature as a variable through the sampling of the
statistical density matrix, implemented by the path integral Monte Carlo
(PIMC) method.~\cite{ceperley}  

QMC methods simulate quantum systems using imaginary-time dynamics since
they are intended for achieving the lowest-energy state. Therefore,
having no access to real-time evolution one looses the possibility of
getting the dynamic structure factor by a Fourier transform of the
intermediate scattering function $F({\bm q},t)$, as it happens in
simulations of classical systems using Molecular Dynamics. Quantum
simulations are able to sample this time-dependent function but in
imaginary time $\tau$, $F({\bm q},\tau)$, and from it to get the dynamic
response through an inverse Laplace transform. But it is well known that
this inverse transform is an ill-posed problem. This means, at the
practical level, that the always finite statistical error of QMC data makes
impossible to find a unique solution for the dynamic structure factor.   

Inverse problems in mathematical physics are a long-standing topic in which
elaborated regularization techniques have been specifically
developed.~\cite{kaipio}
Focusing on the inversion of QMC data, to extract the dynamic response,
several methods have been proposed in the last years. Probably, the most
used approach is the Maximum Entropy (ME) method which incorporates some
a priori expected behavior through an entropic term.~\cite{jarrell} This method works
quite well if the response is smooth but it is not able to reproduce
responses with well-defined peaks. In this respect, other methods have
recently proved to be more efficient than ME. For instance, the average
spectrum method (ASM),~\cite{sandwik} the stochastic optimization method
(SOM),~\cite{mischenko} the method
of consistent constraints (MCC),~\cite{svistunov} and the genetic inversion via falsification
of  theories (GIFT) method~\cite{galli} have been able to recover sharp features in
$S({\bm q},\omega)$ which ME smoothed out. All those methods are
essentially stochastic optimization methods using different strategies and
constraints. It is also possible to work out the inverse problem without 
stochastic grounds~\cite{rota} by using the Moore-Penrose pseudoinverse and a Tikhonov
regularization.~\cite{tikhonov} Other approaches try to reduce the ill-conditioned
character of this inverse problem by changing the kernel from the Laplace
transform to a Lorentz one.~\cite{pederiva} Finally, the computation of complex-time
correlation functions has been recently realized in simple problems and
proved to be able to severely reduce the ill-nature of the Laplace
transform.~\cite{rota} 
   
In this paper, we use the PIMC method to estimate the dynamic response of
liquid $^4$He in a range of temperatures covering the normal-superfluid
transition at $T_\lambda= 2.17$ K. The inversion method from imaginary time
to energy is carried out via the simulated annealing method, which is 
a well-known stochastic multidimensional optimization method widely used in
physics and engineering.~\cite{annealing} Our method is rather similar to the GIFT
one~\cite{galli} but
changing the genetic algorithm by simulated annealing. 
The GIFT method was applied to the study of the dynamic response of liquid
$^4$He at zero temperature and proved to work much better than ME,
producing a rather sharp quasi-particle peak and also some structure at
large energies, corresponding to multiparticle excitations.~\cite{galli} 
The temperature
dependence of $S({\bm q},\omega)$ has been much less studied. 
Apart from a quantum-semiclassical estimation of the response at high
$q$,~\cite{nakayama} the only
reported results where obtained by combining PIMC and the ME method which
worked well in the normal phase but not in the superfluid
part.~\cite{bonin} Therefore,
the significant effect of the temperature on the dynamics of the liquid
through the $\lambda$ transition was lost. We show that the improvement on
the inversion method leads to a significantly better description of 
$S({\bm q},\omega)$ in all the temperature range studied, with reasonable
agreement with experimental data.

The rest of the paper is organized as follows. A short description of the
PIMC method and a discussion of the inversion method used is contained in
Sec. II. In Sec. III, we report the results achieved for the dynamic
response, excitation spectrum, phonon strength and lowest energy-weighted
sum rules across
the transition. Finally, the main conclusions and a summary of the main
results are contained in Sec. IV.

\section{Method}
\label{sec:methods}

The thermal density matrix of a quantum system is given by
\begin{equation}
\hat{\rho} = \frac{e^{-\beta \hat{H}}}{Z} \ ,
\label{rho}
\end{equation}
where $\beta=1/(k_B T)$, $k_B$ is the Boltzmann constant, 
and $Z=\text{Tr} (e^{-\beta \hat{H}})$ is the partition
function. The knowledge of $\hat{\rho}$ allows for the calculation of the 
expected value of any operator $\hat{O}$,
\begin{equation}
\langle \hat{O} \rangle = \text{Tr} (\hat{\rho} \, \hat{O}) \ ,
\label{expected}
\end{equation} 
which in coordinate representation turns to
\begin{equation}
\langle \hat{O} \rangle = \int d \bm{R} \ \rho(\bm{R},\bm{R};\beta) \,
O(\bm{R}) \  , 
\label{expectedr}
\end{equation}     
with $\bm{R}=\{ \bm{r}_1,\ldots,\bm{r}_N\}$ for an $N$-particle system.
Deep in the quantum regime, i.e. at very low temperature, the estimation of
the density matrix for a many-body system is obviously a hard problem.
However, the convolution property of $\hat{\rho}$,
\begin{equation}
\rho(\bm{R}_1,\bm{R}_{M+1};\beta) = \int d \bm{R}_2 \ldots d \bm{R}_M \
\prod_{j=1}^{M} \rho(\bm{R}_j,\bm{R}_{j+1};\tau) \ , 
\label{convolution}
\end{equation}
with $M$ an integer and $\tau=\beta/M$, shows how to build the density
matrix at the desired temperature $T$ from a product of density matrices at a
higher temperature $MT$. If the temperature is large enough, one is able to
write accurate approximations for $\hat{\rho}$ and thus the quantum density
matrix can be calculated, as stated by the Trotter formula
\begin{equation}
e^{-\beta (\hat{K} + \hat{V})} = \lim_{M \to \infty} \left( e^{-\tau
\hat{K}} e^{-\tau \hat{V}} \right)^M \ . 
\label{trotter}
\end{equation} 
In Eq. (\ref{trotter}), we have considered a Hamiltonian $\hat{H}=\hat{K} +
\hat{V}$, with $\hat{K}$ and $\hat{V}$ the kinetic and potential operators,
respectively. In the limit of high temperature the system approaches the
classical regime where  $e^{-\beta (\hat{K} + \hat{V})} = 
e^{-\beta \hat{K}} e^{-\beta \hat{V}}$. This factorization, called
primitive approximation, is however not accurate enough to simulate a
quantum liquid as $^4$He because the number of required terms (\textit{beads}) $M$
is too large.  To make  our PIMC simulations of superfluid $^4$He reliable, we
use a fourth-order time-step ($\tau$) approximation due to Chin, following
the implementation discussed in Ref. \onlinecite{kostas}. Liquid $^4$He is
a Bose liquid and thus we need to sample not only particle positions but
permutations among them. To this end, we use the worm
algorithm.~\cite{worm}

In the present work, we are mainly interested in calculating the
intermediate scattering function $F(\bm{q},\tau)$, defined as
\begin{equation}
F(\bm{q},\tau) = \frac{1}{N} \, \langle \hat{\rho}_{\bm{q}}(\tau)
\, \hat{\rho}_{\bm{q}}^\dagger(0) \rangle \ ,
\label{intermediate}
\end{equation}
with $\hat{\rho}_{\bm{q}}(\tau)=\sum_{i=1}^N e^{i \bm{q} \cdot \bm{r}_i}$ the
density fluctuation operator. The function $F(\bm{q},\tau)$ is the Laplace
transform of the dynamic structure factor $S(\bm{q},\omega)$ which
satisfies the detailed balance condition,
\begin{equation}
S(\bm{q},-\omega) = e^{-\beta \omega} S(\bm{q},\omega) \ ,
\label{balance}
\end{equation}
relating the response for negative and positive energy transfers
$\omega$. Taking into account Eq. (\ref{balance}), one gets
\begin{equation}
F(\bm{q},\tau) = \int_0^\infty d \omega \  S(\bm{q},\omega) (e^{-\omega \tau}
+ e^{-\omega(\beta-\tau)}) \ .
\label{laplace}
\end{equation}
The intermediate scattering function is periodic with $\tau$, as it can be
immediately seen from Eq. (\ref{laplace}): $F({\bm
q},\beta-\tau)=F(\bm{q},\tau)$. Therefore, it is necessary to sample this
function only up to $\beta/2$ (half of the polymer representing each particle in
PIMC terminology). From the PIMC simulation, one samples $F(\bm{q},\tau)$
at the discrete points in which the action at temperature $T$ is
decomposed (Eq. \ref{convolution}).

\begin{figure}
\begin{center}
\includegraphics[width=0.9\linewidth,angle=0]{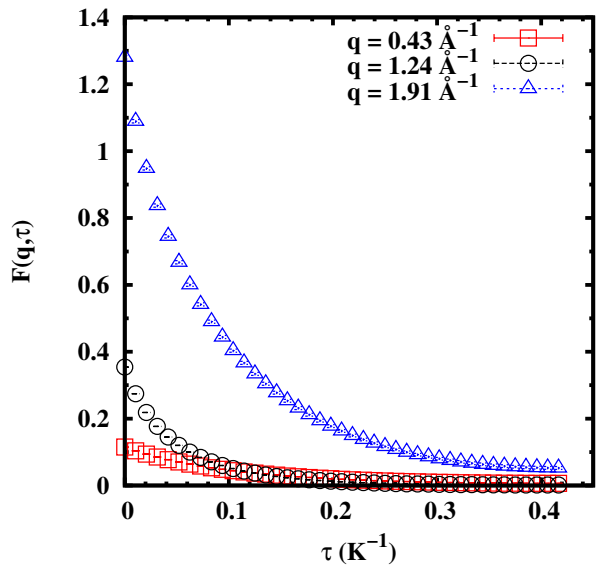}
\caption{(Color online) Intermediate scattering function computed for
$^4$He at saturated vapor pressure 
($\rho = 0.021858 \, \textrm{\AA}^{-3}$) and $T = 1.2$ K, for different values of $q$.
}
\label{Fig:1}
\end{center}
\end{figure}

In Fig. \ref{Fig:1}, we show the characteristic behavior of
$F(\bm{q},\tau)$ for three different $q$ values at $T=1.2$ K. These
are monotonously decreasing functions ending at a finite value at $T/2$
which approaches zero when $T \to 0$. The initial point at $\tau=0$
corresponds to the zero energy-weighted sum rule of the dynamic response,
which in turn is the static structure factor at that specific $q$ value,
\begin{equation}
m_0 = S(\bm{q}) = \int_{-\infty}^{\infty} d \omega \  S(\bm{q},\omega) \ .
\label{sq}
\end{equation}
With the PIMC results for $F(\bm{q},\tau)$, the next step is to find a
reasonable model for $S(\bm{q},\omega)$ having always in mind the
ill-conditioned nature of this goal. In our scheme, we assume a step-wise
function,
\begin{equation}
S_{\text m}(q,\omega) = \sum_{i=1}^{N_s} \xi_i \ \Theta(\omega-\omega_i)
\, \Theta(\omega_{i+1}-\omega) \ ,
\label{sqwmodel}
\end{equation}
with $\Theta(x)$ the Heaviside step function, and $\xi_i$ and $N_s$
parameters of the model. As our interest relies on the study of homogeneous
translationally invariant systems, the response functions depend only of the
modulus $q$. Introducing $S_{\text m}(q,\omega)$ in Eq.
(\ref{laplace}), one obtains the corresponding model for the intermediate
scattering function,
\begin{eqnarray}
F_{\text m}(q,\tau)  =  \sum_{i=1}^{N_s} & \xi_i & \ \left[ \frac{1}{\tau}
\left( e^{-\tau \omega_i} - e^{-\tau \omega_{i+1}} \right) \right.
\label{fmodel} \\
&&  \left. + \frac{1}{\beta-\tau} \left( e^{-(\beta-\tau)\omega_i} -
e^{-(\beta-\tau) \omega_{i+1}} \right) \right] \nonumber 
\end{eqnarray}

\begin{figure}
\begin{center}
\includegraphics[width=0.95\linewidth,angle=0]{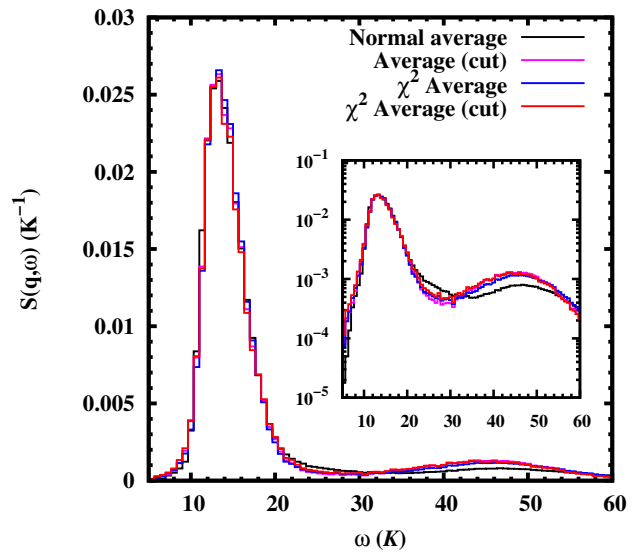}
\caption{(Color online) Dynamic structure factor at $T=1.2 $ K, 
saturated vapor pressure (SVP),
and $q = 0.76 \, \textrm{\AA}^{-1}$ using different averaging methods.
Inset shows same data but using a log scale in the $y$-axis.
}
\label{Fig:2}
\end{center}
\end{figure}

Written in this way, the inverse problem is converted into a multivariate
optimization problem which tries to reproduce the PIMC data with the
proposed model, Eq. \ref{fmodel}. To this end, we use the simulated annealing
method which relies on a thermodynamic equilibration procedure from high to
low temperature according to a predefined template
schedule.~\cite{annealing} The cost function to be minimized is the
quadratic dispersion,
\begin{equation}
\chi^2(q) = \sum_{i=1}^{N_p} \left[ F(q,\tau_i) - F_{\text m}(q,\tau_i)
\right]^2 \ ,
\label{chi2}
\end{equation} 
with $N_p$ the number of points in which the PIMC estimation of the
intermediate scattering function is sampled. Eventually, one can also
introduce as a denominator of Eq. (\ref{chi2}) the statistical errors
coming from the PIMC simulations. However, we have checked that this is not
affecting so much the final result since the size of the errors is rather
independent of $\tau$.

\begin{figure}
\begin{center}
\includegraphics[width=0.45\textwidth,angle=0]{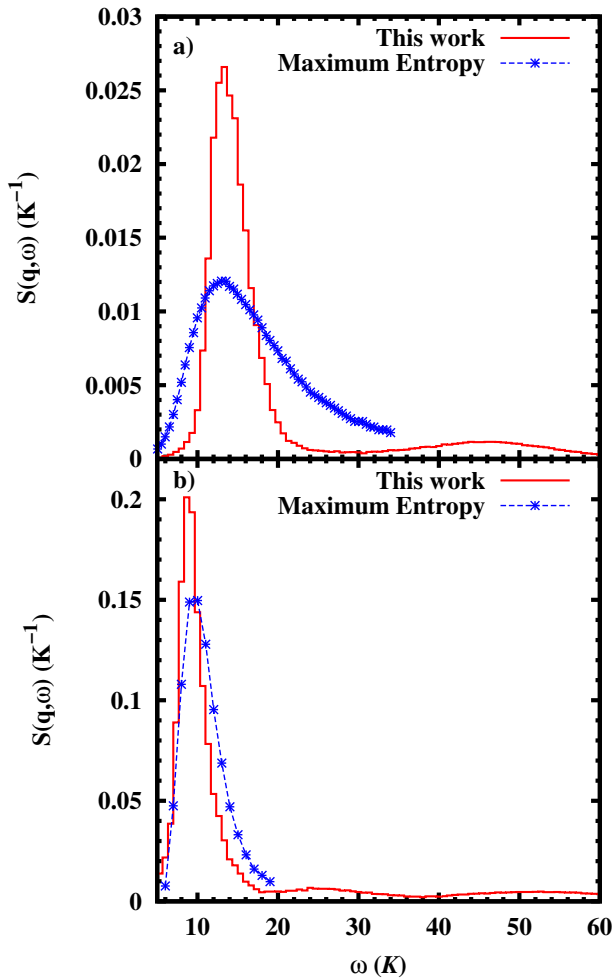}
\caption{(Color online) Comparison between the present results for the dynamic
structure factor and those obtained in Ref. \onlinecite{bonin} using the maximum entropy
method for $q=0.76$ \AA$^{-1}$(a) and $1.81$ \AA$^{-1}$ (b). Both results are calculated at
SVP and $T=1.2$ K. }
\label{Fig:3}
\end{center}
\end{figure}

The optimization
leading to $S(q,\omega)$ is carried out over a number $N_t$ of independent
PIMC calculations of $F(q,\tau)$. Typically, we work with a population 
$N_t=24$ and for each one we perform a number $N_a=100$ of independent simulated 
annealing searches. The mean average of these $N_a$ optimizations is our
prediction for the dynamic response for a given $F(q,\tau)$. We also
register the mean value of $\chi^2$ (Eq. \ref{chi2}) of the $N_a$
optimizations. As an example, the mean value of $\chi^2$ in a
simulation with data at $T=1.2$ K and  $q=1.91$ \AA$^{-1}$ 
is $2.19 \cdot 10^{-5}$, with minimum and maximum values of 
$2.37 \cdot 10^{-6}$ and $3.80 \cdot 10^{-4}$, respectively.
At this same temperature, $N_p=41$ and the number of points of the model
$S(q,\omega)$ (Eq. \ref{fmodel}) is $N_s=150$.

With the outcome for the $N_t$ series we have tried
different alternatives to get the final prediction. We can take just the
statistical mean of the series or a weighted mean, in which the weight of
each function is the inverse of its corresponding $\chi^2$, to give more
relevance to the best-fitted models. Additionally, we have also tried to
make both of these estimations but selecting the 20\% best functions
according to its $\chi^2$. In Fig. \ref{Fig:2}, we plot the results
obtained following these different possibilities. All the results are quite
similar, with minor differences; only at large energies we can observe that
the weighted mean gives slightly more structure (see inset in Fig.
\ref{Fig:2}). Also, the effect of
selecting the best $\chi^2$ models seems to be not much relevant.

In Fig. \ref{Fig:3}, we compare the results obtained for the dynamic
structure factor at $T=1.2$ K and saturated vapor pressure (SVP) with previous 
results obtained using the Maximum Entropy (ME) method.~\cite{bonin} The ME
results are significantly broader, mainly at the lowest $q$ value, and with
only smooth features. This broadening is probably a result of the entropic
prior used in those estimations, which seems to favor smooth solutions. In
the figure, we can observe that the position of the ME peak is coincident with
ours but the ME solution lacks of any structure beyond the quasi-particle
peak (see Appendix for additional comparisons between ME and our stochastic
optimization procedure). In our estimation, we do not use any prior information in the search
of optimal reconstructions and thus it is free from any a priori
information except that the function is positive definite for any energy.
Moreover, the simulated annealing optimization leads to dynamic responses
that fulfil the energy-weighted sum rules $m_0$ and $m_1$,
\begin{equation}
m_1= \int_{-\infty}^{\infty} d \omega \ \omega S(q,\omega) = \frac{\hbar^2
q^2}{2m} \ ,
\label{m1}
\end{equation}
without imposing them as constraints in the cost
function (Eq. \ref{chi2}). Also, the $m_{-1}$ sum-rule, related to the static
response, is in agreement with experiment (see next Section).

\section{Results}
\label{sec:results}

We have performed PIMC calculations of liquid $^4$He following the SVP
densities, from $T=0.8$ to $4$ K. The interatomic potential is of Aziz
type~\cite{aziz} and the number of particles in the simulation box, under
periodic boundary conditions, is $N=64$. In some cases we have used a
larger number of particles ($N=128$) without observing any significant change in $F(q,\tau)$.
The number of terms $M$ (Eq. \ref{convolution}) is large enough to eliminate
any bias coming from the path discretization; we used $\tau=0.0104$
K$^{-1}$.

\begin{figure}
\begin{center}
\includegraphics[width=0.90\linewidth,angle=0]{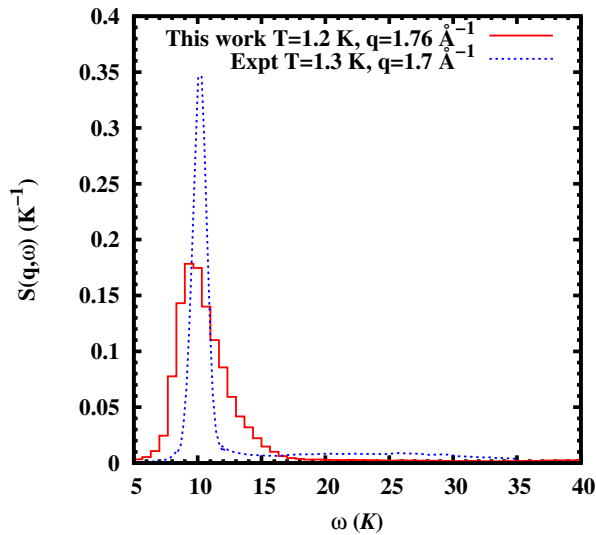}
\caption{(Color online) Dynamic structure factor at $T =
1.2$ K and $q = 1.76 \, \textrm{\AA}^{-1}$ 
compared with experimental data ($T=1.3K$, $q = 1.7 \,
\textrm{\AA}^{-1}$).~\cite{andersen}
}
\label{Fig:4}
\end{center}
\end{figure}

\begin{figure}
\begin{center}
\includegraphics[width=0.9\linewidth,angle=0]{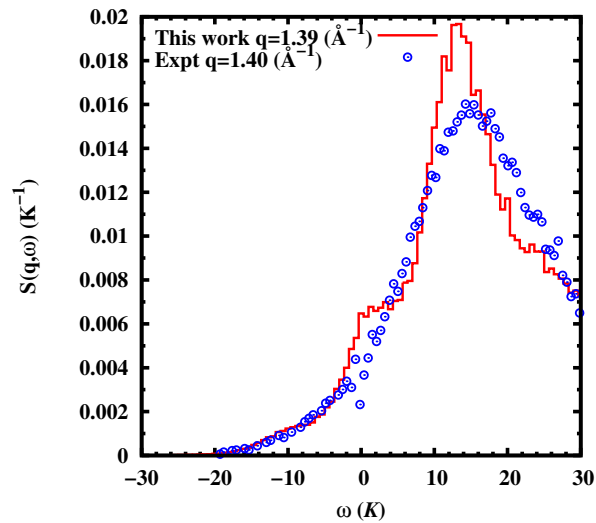}
\caption{(Color online) Dynamic structure factor at $T = 4.0$ K and 
$q = 1.40 \, \textrm{\AA}^{-1}$. The experimental data is  
from Ref. \onlinecite{andersen-bonin}.
}
\label{Fig:5}
\end{center}
\end{figure}

We compare our result for the dynamic response in the superfluid phase with
experimental data from Ref. \onlinecite{andersen} in Fig. \ref{Fig:4}. The
theoretical peak is located around an energy which is very close to the
experimental one but it is still broader than in the experiment. However,
the strength (area) of this peak is in good agreement with the experimental
one, as we will comment later. The quasi-particle peak disappears in the
normal phase, above $T_\lambda$, as we can see in Fig. \ref{Fig:5}. In this
figure, we compare our results at $T=4$ K with experimental outcomes at the
same $T$. In this case, we see that both the position of the peak and its
shape is in an overall agreement with the experiment.

\begin{figure}
\begin{center}
\includegraphics[width=0.9\linewidth,angle=0]{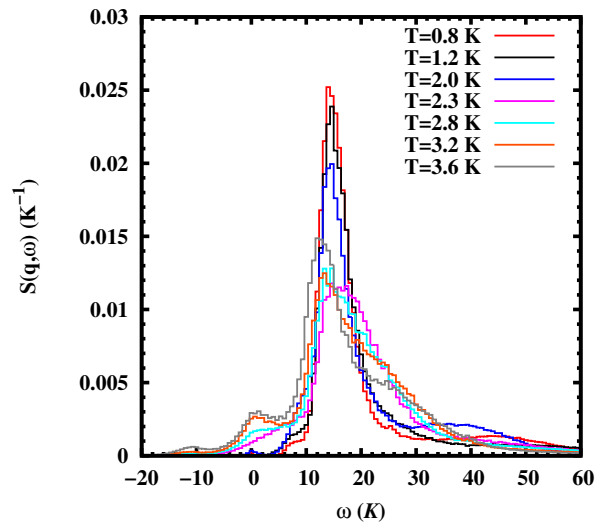}
\caption{(Color online) Dynamic structure factor of liquid $^4$He 
for $q = 0.88 \, \textrm{\AA}^{-1}$ at different temperatures.
}
\label{Fig:6}
\end{center}
\end{figure}

One of the main goals of our study has been the study of the effect of the
temperature on the dynamics of liquid $^4$He. In Fig. \ref{Fig:6}, we
report results of $S(q,\omega)$ in a range of temperatures from $T=0.8$ to
$4$ K in the phonon region of the spectrum, with $q=0.88$ \AA$^{-1}$. At
this low $q$ value, the behavior with $T$ is not much different for the
superfluid and normal phases, a feature which is also observed in neutron
scattering data.~\cite{glyde} We observe a progressive broadening of the
peak with $T$ which appears already below $T_\lambda$ and continues above
it. Even at the highest temperature $T=4$ K, we identify a collective peak
corresponding to a sound excitation.~\cite{glyde} The main difference between both
regimes is that the quasi-particle energy below $T_\lambda$ is nearly
independent of $T$ whereas, in the normal phase, this energy decreases in
a monotonous way.

\begin{figure}
\begin{center}
\includegraphics[width=0.9\linewidth,angle=0]{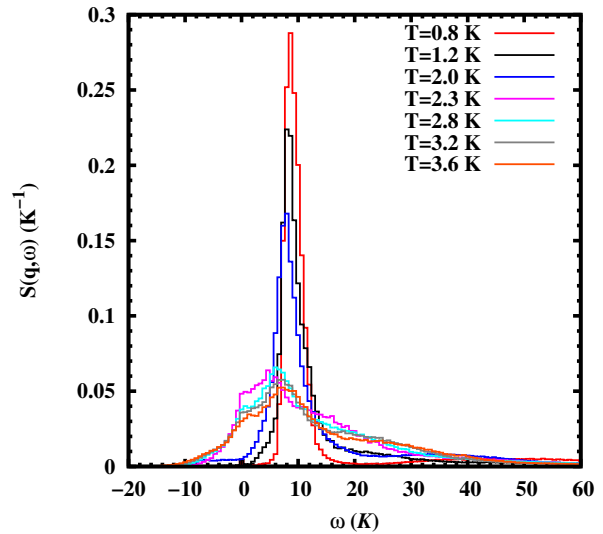}
\caption{(Color online) Dynamic structure factor of liquid $^4$He 
for $q = 1.91 \, \textrm{\AA}^{-1}$ at different temperatures.}
\label{Fig:7}
\end{center}
\end{figure}

Near the roton, the dependence of the dynamic response with $T$ is
significantly different. In Fig. \ref{Fig:7}, we report results of
$S(q,\omega)$ at $q=1.91$ \AA$^{-1}$  at different temperatures across
$T_\lambda$. The most relevant feature is the drop of the quasi-particle
peak for $T>T_\lambda$. In the superfluid phase, the peak remains sharp
with a nearly constant energy. Just crossing the transition (in our data
for $T\ge 2.3$ K), the peak disappears and only a broad response is
observed, with an energy that moves slightly down. According to the Landau
criterium the existence of a roton gap implies a  critical velocity larger
than zero and thus a superfluid phase. Our PIMC data is consistent with this
picture since we observe as the resulting superfluid density, derived from the winding
number estimator, goes to zero at $T_\lambda$, in agreement with 
the disappearance of the roton excitation in $S(q,\omega)$.

\begin{figure}
\begin{center}
\includegraphics[width=0.9\linewidth,angle=0]{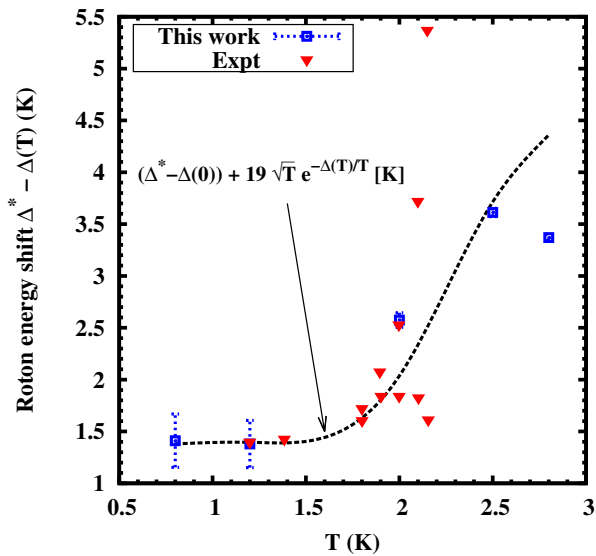}
\caption{(Color online) Temperature dependence of the roton energy.
The experimental points and suggested fit are from Ref.
~\onlinecite{rotonshift} with $\Delta(0) = 8.62$ K . In the fit, 
$\Delta^* > \Delta(T)$ stands for an  arbitrary energy value.
}
\label{Fig:8}
\end{center}
\end{figure}

Our results for the temperature dependence of the roton energy $\Delta(T)$ 
are shown in Fig. \ref{Fig:8}. For temperatures $T < 1.5$ K, $\Delta(T)$ is
practically constant around a value $8.60$ K, in agreement with
experiment.~\cite{rotonshift}. For larger temperatures, still in the
superfluid part, this energy gap starts to decrease with the largest change
around the transition temperature. For temperatures $T> 2.5$ K, the peak
vanishes and $\Delta(T)$ flattens but then one really can not continue
speaking about the roton mode. In the same figure we report experimental
results for the roton energy in the superfluid phase. At same temperature,
our results agree well with the experimental ones which show some erratic
behavior around $T \simeq 2$ K but compatible with a decrease of the roton
gap with $T$. Still in the same figure, we report the fit used in Ref.
\onlinecite{rotonshift}, that is based on the roton-roton interaction
derived from Landau and Khalatnikov theory.~\cite{landau} This law seems to
be right only at the qualitative level, with significant deviation with our
results and still larger discrepancies with the experimental values.

\begin{figure}
\begin{center}
\includegraphics[width=0.9\linewidth,angle=0]{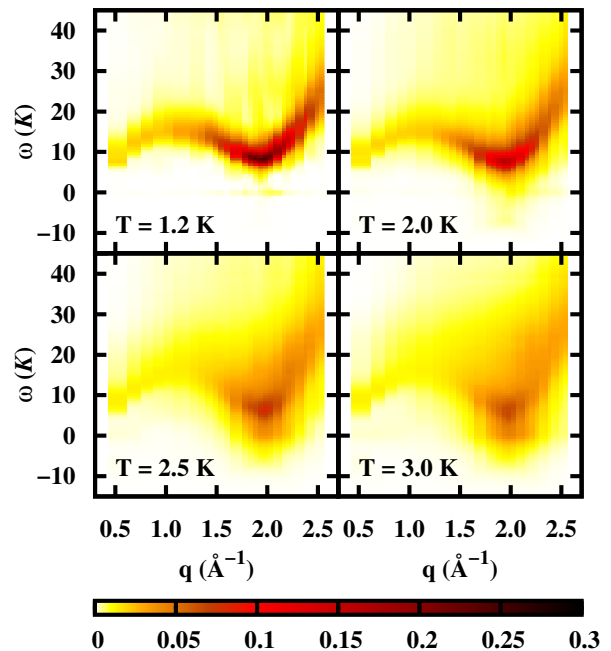}
\caption{(Color online) Color map of the dynamic response in the
momentum-energy plane at different temperatures, below and upper
$T_\lambda$.
}
\label{Fig:9}
\end{center}
\end{figure}

The results obtained for $S(q,\omega)$ in the present calculation are
summarized in Fig. \ref{Fig:9} as a color map in the momentum-energy plane.
In the superfluid phase, the phonon-roton curve is clearly observed, with
the highest strength of the quasi-particle peak located in the roton
minimum, in agreement with experiment. The multiparticle part above the
single-mode peak is also observed but without any particular structure. At
$T=2$ K the roton peak is still observed but some intensity starts to
appear below it, At $T=2.5$ and 3 K, we still obtain intensity in the roton but
the peak, and in general, all the spectrum appears much more diffuse.

\begin{figure}
\begin{center}
\includegraphics[width=0.9\linewidth,angle=0]{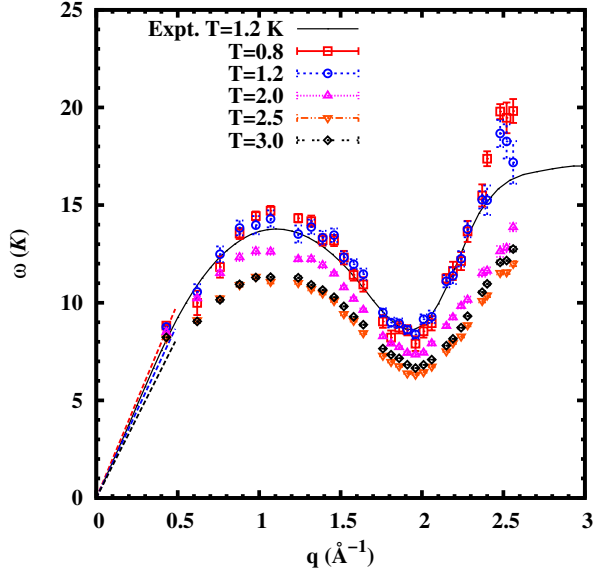}
\caption{(Color online) Phonon-roton spectrum of liquid $^4$He at 
different temperatures. The line corresponds to experimental data at
$T=1.2$ K.~\cite{cowley1,cowley2}. Straight lines at small $q$ stand for
the low $q$ behavior, $\omega=c q$ with $c$ the speed of sound, at
temperatures $T=0.8$, $1.2$, and $3.0$ K (from larger to smaller slope).
}
\label{Fig:10}
\end{center}
\end{figure}

The excitation energy of the collective mode is shown in Fig. \ref{Fig:10}
at different temperatures. Our results at the lowest temperatures, 
$T=0.8$ and $1.2$ K, are indistinguishable within the statistical errors
and are in close agreement with the inelastic neutron scattering data at
$T=1.2$ K from
Refs. \onlinecite{cowley1,cowley2}, except at the end of the spectrum
(Pitaevskii plateau). In fact, for $q>2.5$ \AA$^{-1}$ the dynamic response
that we obtain from the reconstruction of the imaginary-time intermediate
scattering function is rather broad and one can not distinguish the double
peak structure observed in experiments. Also, notice that the energies
corresponding to $q \alt 0.5$ \AA$^{-1}$ are not accessible in our
simulations since our minimum $q_{\text{min}}$ value is restricted to be $2
\pi/L$, with $L$ the length of the simulation box. At $T=2$ K, very close
to the superfluid transition temperature, we observe as the energies of the
maxon and roton modes significantly decrease whereas the phonon part is
less changed. When the temperature is above the transition, we can observe
that the maximum of the peaks, now much broader, seem to collapse again in
a common curve  around the maxon. Instead, in the roton it seems that the
energy could increase again at the largest temperature. This latter feature
is quite unexpected and could be a result of our difficulty in the localization
of the maximum in a rather broad dynamic response. The overall description
on the evolution of the phonon-roton spectrum with $T$ is in close
agreement with experimental data.~\cite{glyde}.

\begin{figure}
\begin{center}
\includegraphics[width=0.9\linewidth,angle=0]{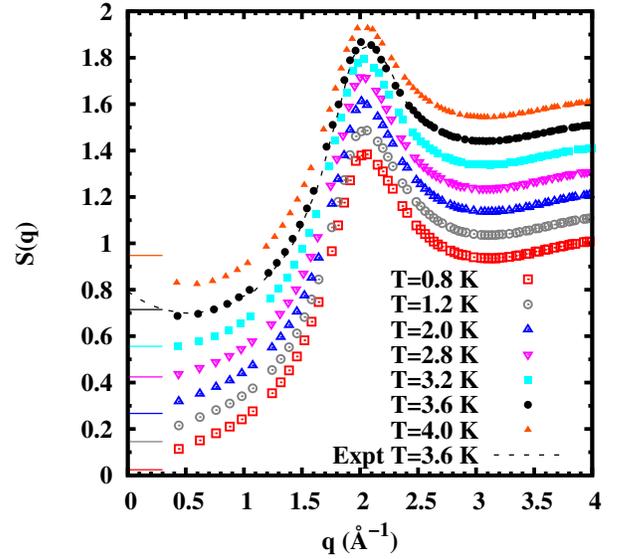}
\caption{(Color online) Static structure factor $S(q)$ at different
temperatures across $T_\lambda$. 
The results have been shifted vertically a constant value to make its
reading easier. The dashed line stands for experimental data from Ref.
\onlinecite{svensson}. Short horizontal lines at $q=0$ correspond to the
value (Eq. \ref{compressibility}) obtained from PIMC.
}
\label{Fig:11}
\end{center}
\end{figure}

The static structure factor $S(q)$ is the zero energy-weighted sum rule of
the dynamic response (Eq. \ref{sq}). This function can be exactly calculated
using the PIMC method as it is the value of the imaginary-time intermediate
scattering function at $\tau=0$. In Fig. \ref{Fig:11}, we show results of
$S(q)$ for the range of analyzed temperatures. The effect of the
temperature on the position and height of the main peak is quite small, in
agreement with the x-ray experimental data from Ref. \onlinecite{svensson}.
We observe a small displacement of the peak to larger $q$ values and a
simultaneous decrease of the height when $T$ increases. These effects can
be mainly associated to the decrease of the density along SVP when the
temperature grows. For values $q \alt 0.5$ \AA$^{-1}$ we do not have
available data due to the finite size of our simulation box. Therefore, we
can not reach the zero momentum value which is related to the isothermal
compressibility $\chi_T$ through the exact relation
\begin{equation}
S(q=0)= \rho k_B T \chi_T \ ,
\label{compressibility}
\end{equation} 
with $k_B$ the Boltzmann constant and $\rho$ the density. The requirement of
this condition produces that $S(q)$ starts to develop a minimum around $q
\simeq 0.5$ \AA$^{-1}$ when $T$ increases. Our results also show this
feature but for larger $T$ ($\sim 3.6$ K) than in experiments ($\sim 3$ K) 
due to our lack of data at low $q$.

\begin{figure}
\begin{center}
\includegraphics[width=0.9\linewidth,angle=0]{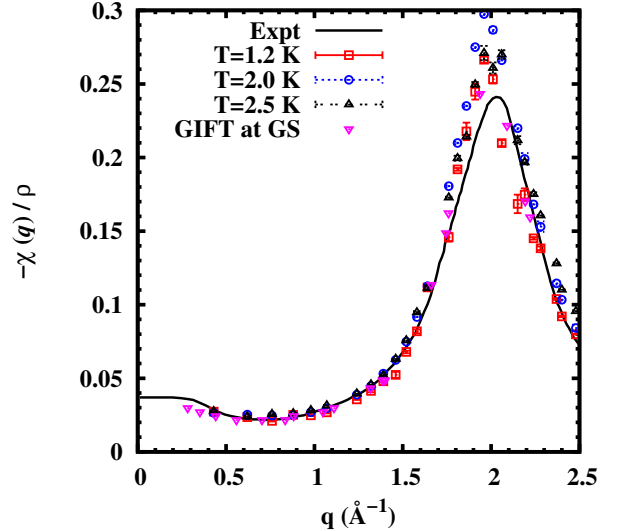}
\caption{(Color online) Static response function at $T = 1.2$,
$2.0$, and $2.5$ K.  For comparison, we also report zero-temperature QMC
results~\cite{galli} and
experimental data obtained at $T=1.2$ K.~\cite{cowley1,cowley2}}
\label{Fig:12}
\end{center}
\end{figure}

From the dynamic structure factor, we can calculate the static response
function $\chi(q)$ since this is directly related to the $1/\omega$ sum
rule through the relation
\begin{equation}
\chi(q)=-2 \rho \int_{-\infty}^{\infty} d \omega \ 
\frac{S(q,\omega)}{\omega} = -2 \rho m_{-1} \ .
\label{chiq}
\end{equation}
The dominant contribution to the $m_{-1}$ sum rule is the quasi-particle
peak and thus it is less sensitive to the multi-phonon part.~\cite{caupin} In Fig.
\ref{Fig:12}, we report the results obtained for $\chi(q)$ at temperatures
$1.2$, $2.0$, and $2.5$ K. We observe that at low $q$ the effect of $T$ is
negligible but around the peak, $q \simeq 2$ \AA$^{-1}$, is really large.
In the superfluid regime, the height of the peak clearly increases with $T$, 
a feature that has not
been reported previously neither from theory nor from experiment. 
At $T=2.5$ K, in the normal phase, the main peak decreases again 
in agreement with the absence of the roton.
In the figure, we plot experimental data~\cite{cowley1,cowley2} at $T=1.2$ K which
is close to our result at low $q$ but with less strength in the peak.
Results from QMC at zero temperature from Ref. \onlinecite{galli} are in an
overall agreement with ours at the lowest $T$, but somehow ours have a
slightly higher peak.

\begin{figure}
\begin{center}
\includegraphics[width=0.9\linewidth,angle=0]{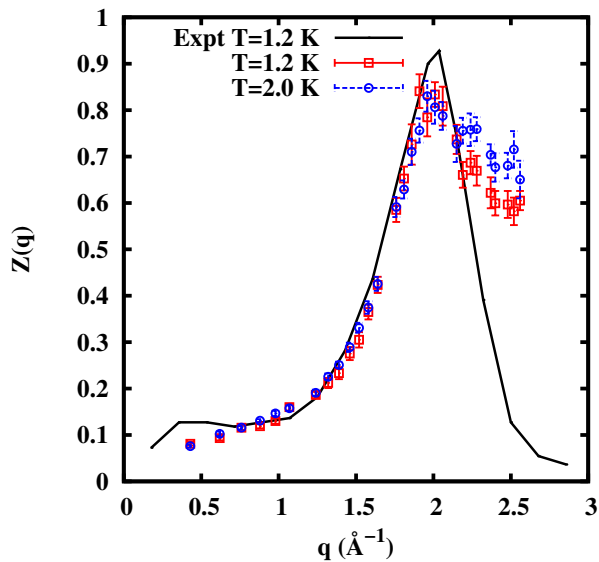}
\caption{(Color online) One-phonon contribution to the dynamic response,
$Z(q)$, at different temperatures. Experimental results 
at $T=1.2$ K from Refs. \onlinecite{cowley1,cowley2}.}
\label{Fig:13}
\end{center}
\end{figure}

The dynamic response of liquid $^4$He is usually written as the sum of two
terms,
\begin{equation}
S(q,\omega) = S_1(q,\omega)+S_{\text{m}}(q,\omega) \ ,
\label{onephonon}
\end{equation}
where $S_1(q,\omega)$ stands for the sharp quasi-particle peak and
$S_{\text{m}}(q,\omega)$  includes the contributions from scattering of more 
than one phonon (multiphonon part). The intensity (area) below the sharp
peak is the function $Z(q)$ which we report in Fig. \ref{Fig:13}. Our
results are compared with experimental data at $T=1.2$ K from Refs.
\onlinecite{cowley1,cowley2}. As we commented previously, our
quasi-particle peaks are less sharp than the experimental ones due to the
uncertainties in the inversion problem from imaginary time to energy.
However, the area below the peak is not so far from the experimental
outcomes. Up to the maximum of the peak, our results are compatible with
the experimental function. However, our data lead to a peak with
less strength and after that, for larger momenta, our results scatter significantly 
due to the
difficulties in the determination of the area below the peak. The
uncertainties in the area estimation do not allow for the observation of an
enhancement of the peak's height when $T$ increases, as reported in
experiments.~\cite{caupin}

\section{Conclusions}
\label{sec:conclusions}

We have carried out PIMC calculations of liquid $^4$He in a wide range of
temperatures across the normal-superfluid transition $T_\lambda$ to
calculate the imaginary-time intermediate scattering function $F(q,\tau)$.
From these functions one can in principle access to the dynamic response
$S(q,\omega)$ through an inverse Laplace transform. But this is an
ill-conditioned problem that can not be solved to deal with a unique
solution. In recent works,~\cite{galli} it has been shown that the use of stochastic
optimization tools can produce results with a richer structure than
previous attempts relying on the maximum entropy method.\cite{jarrell} We have adopted
here the well-known simulated annealing technique to extract the dynamic
response, without any a priori bias in the search in order to get a result
as unbiased as possible. In spite of the lack of any constraint in the cost
function, we have verified that the three lowest energy-weighted sum rules are 
satisfactorily satisfied giving us some confidence on the reliability of
our algorithm.

The results of the dynamic response are still not enough sharp in the
quasi-particle peaks of the superfluid phase but the position of the peaks
and the area below them are in nice agreement with experimental data. 
Interestingly, our results show clearly the signature of the transition in
the roton peak, whose amplitude drops rapidly for $T>T_\lambda$. The effect of
the temperature on the phonon-roton spectrum, static structure factor, and
static response has been also studied.

The difficulties of extending correlated perturbative approaches to finite
$T$ have lead to a really unexplored dynamics of superfluid liquid $^4$He,
at least from a microscopic approach. With the present work, which can be
considered an extension and improvement of a previous work based on the
maximum entropy method,~\cite{bonin} we have shown that the combination of PIMC and
stochastic reconstruction is able to produce a satisfactory description of
the quantum dynamics at finite temperature. We are also convinced that
in the near future we can improve even more the present results. In this
respect, one of the more promising avenues could be the estimation of
complex-time correlation functions, instead of the merely imaginary ones, which
can reduce the ill-posed character of the inversion problem due to its
non-monotonic structure.~\cite{rota}

\begin{figure}
\begin{center}
\includegraphics[width=0.9\linewidth,angle=0]{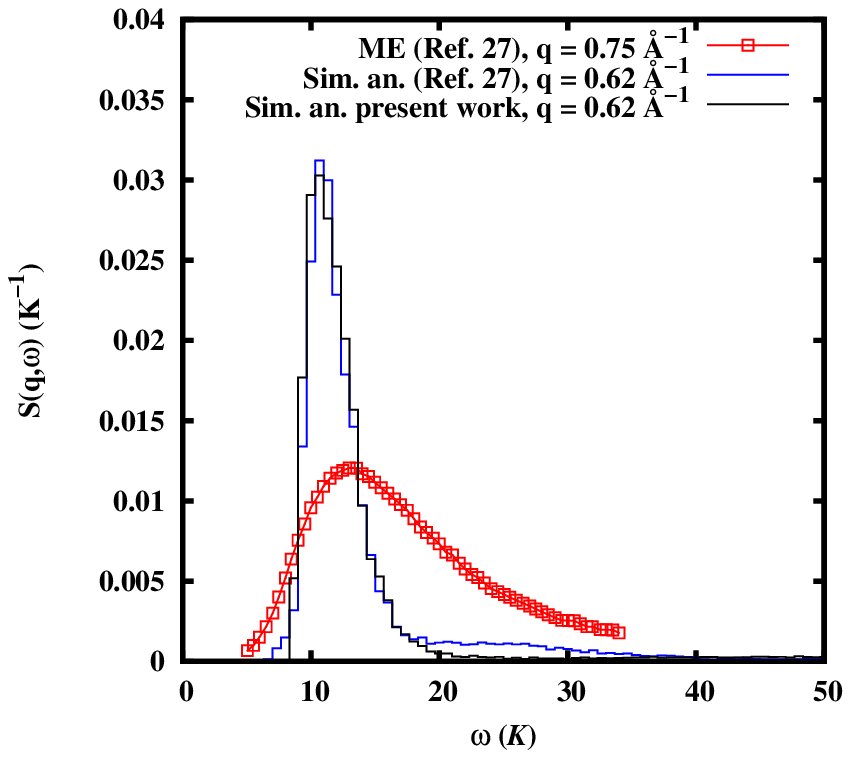}
\caption{(Color online) Comparison between the dynamic response obtained
with ME  and our stochastic optimization method using intermediate
scattering data from Ref. \onlinecite{bonin}.}
\label{Fig:A1}
\end{center}
\end{figure}

\begin{figure}
\begin{center}
\includegraphics[width=0.9\linewidth,angle=0]{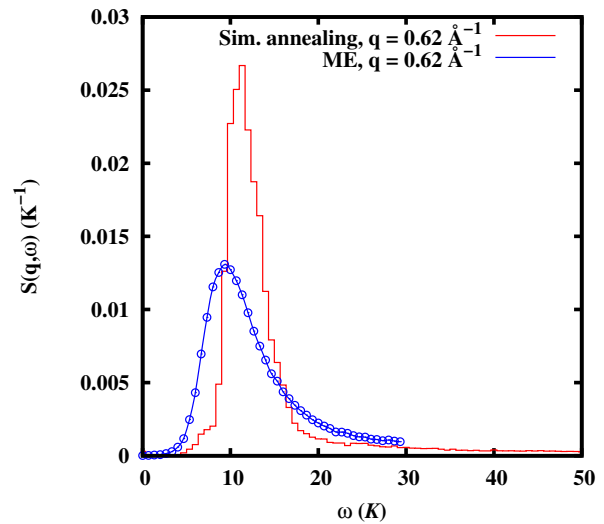}
\caption{(Color online) Comparison between the dynamic response obtained
with ME  and our stochastic optimization method using our intermediate
scattering data.}
\label{Fig:A2}
\end{center}
\end{figure}

\begin{acknowledgments}
This research was supported by MICINN-Spain Grant No. FIS2014-56257-C2-1-P. 
\end{acknowledgments}

\section*{appendix}
In Fig. \ref{Fig:3}, we have compared results for $S(q,\omega)$ derived
from our stochastic optimization method and results reported in Ref.
\onlinecite{bonin} using ME. As the intermediate scattering
function $F(q,\tau)$ used in both estimations is different and used by
different authors it could happen that the differences observed in Fig.
\ref{Fig:3} were due more to the differences between the calculated
imaginary-time response than to the inversion method itself. To clarify
this point, we report in this Appendix results of two additinal
comparisons.

In Fig. \ref{Fig:A1}, we report results for $S(q,\omega)$  at $q=0.62$
\AA$^{-1}$ using our imaginary-time data and stochastic optimization. In
the figure, we also show the dynamic response that we have obtained by
applying our inversion method to the imaginary-time data reported in Ref.
\onlinecite{bonin}. Finally, the figure also shows the ME results reported
in Ref. \onlinecite{bonin} but for a slightly different $q$ value since
results for $q=0.62$ \AA$^{-1}$  are not given in that paper. As one can
see, starting from their published data and applying our method the
results compare favorably with our response $S(q,\omega)$. Therefore,  
the different quality of the input data is so small that no effect is
observed.

In order to make a more clear comparison between both inversion methods we
show in Fig. \ref{Fig:A2} results for the dynamic response using our data
for $F(q,\tau)$. At the same $q$ value than in Fig. \ref{Fig:A1}, we report results 
obtainded with
stochastic optimization and using the ME method. The results are
similar to the ones shown in Fig. \ref{Fig:A1} and lead to the same
conclusion, that is, the ME method generates smoother functions than our
method. This conclusion is in agreement with a similar analysis reported by
Vitali \textit{et al.}~\cite{galli}.

\end{document}